\documentclass[aps,prb,twocolumn,groupedaddress,floats,showpacs]{revtex4}
\usepackage{latexsym}
\usepackage{dcolumn}
\usepackage[dvips]{graphicx}
\usepackage{amssymb}
\usepackage{graphics}
\usepackage{amsmath}
\usepackage{epsf}

\newcommand{\vk}{\ensuremath{{\bf k}}}
\newcommand{\vv}{\ensuremath{{\bf v}}}

\newcommand{\ve}    {\varepsilon}
\newcommand{\beq}    {\begin{equation}}
\newcommand{\enq}    {\end{equation}}

\newcommand{\qq} {\ensuremath{{\bf q}}}
\newcommand{\bqa} {\begin{eqnarray}}
\newcommand{\eqa} {\end{eqnarray}}
\newcommand{\EE} {\ensuremath{{\bf E}}}
\newcommand{\jj} {\ensuremath{{\bf j}}}
\newcommand{\Ical}{\ensuremath{{\cal I}}}
\newcommand{\kp} {\ensuremath{{\bf k'}}}

\begin{document}
\title{Coulomb drag in monolayer and bilayer graphene}
\author{E. H. Hwang, Rajdeep Sensarma, and S. Das Sarma} 
\affiliation{Condensed Matter Theory Center, Department of 
        Physics, University of Maryland, College Park, MD 20742-4111
        U.S.A. }

\begin{abstract}
  We theoretically calculate the interaction-induced frictional Coulomb drag resistivity
  between two graphene monolayers as well as between two graphene
  bilayers, which are spatially separated by a distance ``$d$''. We
  show that the drag resistivity between graphene monolayers can be
  significantly affected by the intralayer momentum-relaxation mechanism. For
  energy independent intralayer scattering, the frictional drag
  induced by inter-layer electron-electron interaction goes
  asymptotically as $\rho_D \sim T^2/n^4d^6$ and $\rho_D \sim
  T^2/n^2d^2$ in the high-density ($k_F d \gg 1$) and low-density
  ($k_F d \ll 1$) limits, respectively. When long-range charge impurity
  scattering dominates within the layer, the monolayer drag
  resistivity behaves as $\rho_D \sim T^2/n^3d^4$ and $T^2 \ln
  (\sqrt{n} d) /n$ for $k_F d \gg 1$ and $k_F d \ll 1$, respectively.
  The density dependence of the bilayer drag is calculated to be
  $\rho_D \propto T^2/n^{3}$ both in the large and small layer
  separation limit. In the large layer separation limit, the bilayer
  drag has a strong $1/d^4$ dependence on layer separation, whereas
  this goes to a weak logarithmic dependence in the strong inter-layer
  correlation limit of small layer separation.
In addition to obtaining the asymptotic analytical formula for Coulomb
drag in graphene, we provide numerical results for arbitrary values of
density and layer separation interpolating smoothly between our
asymptotic theoretical results. 
\end{abstract}
\maketitle

\section {introduction}

Much attention has recently focused on multilayer systems in graphene,
where carrier transport properties may be strongly affected by
interlayer interaction effects
\cite{dassarma:2011,kim:2011,feldman:2009,min:2008,hwang:2009,nand:2010}. In
particular, temperature and density dependent Coulomb drag properties
have recently been studied in the spatially separated double layer
graphene systems \cite{kim:2011}.  Frictional drag measurements of
transresistivity in double layer systems have led to significant
advances in our understanding of density and temperature dependence of
electron-electron interactions in semiconductor-heterostructure-based
parabolic 2D systems
\cite{gramila:1991,zheng:1993,flensberg:1995,rojo:1999,hwang:2003,dassarma:2005}.
While electron-electron interactions have indirect consequences (for
example, through carrier screening) for
transport properties of a single isolated sheet of monolayer or bilayer
graphene, the Coulomb drag effect provides an opportunity to directly
measure the effects of electron-electron interactions through a
transport measurement, where momentum is transferred from one layer to
the other layer due to inter-layer Coulomb scattering. In Coulomb
drag measurements the role of electron interaction effects can be
controlled by varying, in a systematic manner, the electron density
($n$), the layer separation ($d$), and the temperature ($T$) since the interlayer electron-electron interaction obviously depends on $n$, $d$ and $T$.
(There is also a rather straightforward dependence of the drag on the
background dielectric constant $\kappa$,  arising trivially through the
interaction coupling constant or equivalently the graphene
fine-structure constant, which we do not discuss explicitly.)

In view of the considerable fundamental significance of the issues
raised by the experimental observations \cite{kim:2011}, we present in
this paper a careful theoretical calculation of frictional drag
resistivity, $\rho_D(T)$, both between monolayer graphene (MLG) sheets
and bilayer graphene (BLG) sheets, using Boltzmann transport equation.
The current work is a generalization of the earlier theoretical work
on graphene drag by Tse {\it et al.} \cite{tse:2007}, where the drag
is calculated within the canonical many-body Fermi liquid theory
assuming an {\it energy independent} intralayer momentum relaxation
time. We would like to note that, although an energy independent
relaxation time captures the effects of both short range and screened
Coulomb impurities in 2D electron gas with parabolic dispersion, it
does not correspond to any disorder model in MLG with its linear
dispersion.
In 
this paper, we generalize the formalism for calculating drag
resistivity by including an arbitrary energy dependent intralayer
scattering mechanism within the linear response Boltzmann equation
description. One new feature of our work is that we consider both MLG
and BLG drag on equal theoretical footing,
providing detailed theoretical drag results for both systems.

We find that, for MLG, the energy-dependent intralayer transport
scattering time is the key to understanding the density dependence of
drag resistivity induced by inter-layer electron-electron
interaction. In the presence of an energy independent scattering time,
the low temperature drag goes asymptotically as $\rho_D \sim
T^2/n^4d^6$ and $\rho_D \sim T^2/n^2d^2$ in the high-density ($k_F d
\gg 1$, where $k_F$ is the Fermi wave vector) and low-density ($k_F d
\ll 1$) limits, respectively. 
These interlayer drag results for energy-independent intralayer
momentum relaxation, however, change qualitatively in the presence of
an energy-dependent intralayer relaxation.
For example, if the intralayer scattering is
dominated by the charged impurities, which is believed to be the
dominant intralayer momentum relaxation mechanism in most 
currently available graphene samples on a substrate
\cite{novoselov:2005,tan:2007,chen:2008}, the MLG drag resistivity
behaves asymptotically as $\rho_D \sim T^2/n^3d^4$ and $T^2 \ln
(\sqrt{n} d) /n$ for $k_F d \gg 1$ and $k_F d \ll 1$, respectively.
In the low temperature regime ($T/T_F \ll 1$), the enhanced phase
space for $q=2k_F$ Coulomb backscattering leads to $\sim T^2\ln(T)$
corrections to the usual $T^2$ dependence of the drag resistivity in
ordinary 2D electron systems, i.e. $\rho_D \propto T^2\ln T$.  We find
that, due to the chirality induced suppression of the $q=2k_F$
backscattering in MLG, these $\sim T^2\ln(T)$ corrections to the drag
resistivity are absent in this system. 

We also investigate the Coulomb drag resistivity for two bilayer
graphene sheets separated by a distance ``d''. The low temperature
behavior of BLG drag follows the usual $T^2$ dependence found in MLG
systems.  Both in the weakly correlated large layer separation limit,
$q_{TF}d \gg 1$, and the strongly correlated small separation limit,
$q_{TF}d \ll 1$ ($q_{TF}$ being the Thomas Fermi screening wavevector
in BLG), the bilayer drag shows an inverse cubic dependence on the
carrier density; i.e. $\rho_D \propto T^2/n^{3}$. In the large layer
separation limit, the BLG drag has a strong $1/d^4$ dependence on layer
separation, whereas this goes to a weak logarithmic dependence in the
strong inter-layer correlation (small layer separation) limit.
The BLG drag, in contrast to MLG drag, does not manifest any
qualitative dependence on the intralayer momentum relaxation mechanism
with both short-range disorder scattering and long-range charged
impurity scattering  producing the same interlayer BLG drag.

The paper is organized as follows. In Sec. II we present the general
formula for drag resistivity in 2D materials from a Boltzmann transport
approach, considering arbitrary momentum dependent scattering
times. The analysis here is for arbitrary 2D chiral electron systems
with two gapless bands, and hence is applicable to both MLG and BLG drag
resistivity. In Sec. III, we study in detail, the temperature, density
and layer separation dependence of the drag resistivity in MLG. We
provide both analytic low temperature asymptotic forms and numerical
results for the MLG drag resistivity, using an energy independent as
well as a linearly energy dependent scattering time. In Sec.IV, we
study the drag resistivity of BLG, focusing both on
analytic asymptotic forms and numerical results, within an energy
independent scattering time approximation. Finally, we conclude 
our study in Sec. IV. with a summary of our work and a comparison of our
results with other works in this field.

\section{Drag resistivity in 2D chiral electron systems}

In this section, we will derive the general formula for drag
resistivity of chiral 2D electron-hole systems. We consider two layers ($a$
and $p$) of the material, which are kept separated by an insulating
barrier of thickness $d$, such that the carriers in the different
layers are only coupled through the Coulomb interaction and there is
no tunneling between the two layers.  In drag experiments, an electric
field $\EE_a$ is applied to layer $a$ (the driven or active layer) and
causes a current density $\jj_a$, which induces the electric field $\EE_p$
in layer $p$ (the dragged or passive layer), where the current $\jj_p$is set
to zero. Then the drag resistivity is defined by $\rho_D =
\EE^\alpha_p/\jj^\alpha_a$, where $\alpha$ is the direction along which the current $\jj_{a}$ flows.~\cite{flensberg:1995,zheng:1993}.

We will consider each layer to be composed of a chiral two-band
(electron and hole) material where the density of the carriers can be
changed independently. Let the dispersion of these bands be given by
$\ve_{s\vk}$, and the corresponding two-component electron wavefunctions be
given by $\psi_{s\vk}$, where $s=\pm 1$ denotes the band indices.The group velocity of the bands are given by $\vv_{s\vk}=\nabla_{\vk}\ve_{s\vk}$.

Let $\tilde{f}^i_{s\vk}$ denote the non-equilibrium distribution function of the band $s$ in the layer $i$, where $(i=a,p)$. The Boltzmann equation for the distribution function is 
\beq
e\EE_i\cdot \vv_{s\vk}\frac{\partial \tilde{f}^i_{s\vk}}{\partial \ve_{s\vk}}={\cal I}^i_{s\vk}
\label{eqn:Boltzmann}
\enq
where, ${\cal I}_{s\vk}$ is the collision integral. Within a
linearized relaxation-time approximation, 
\beq
\tilde{f}^i_{s\vk}=f^i_{s\vk}-e\tau^i_\vk\EE_i\cdot \vv_{s\vk}\frac{\partial f^i_{s\vk}}{\partial \ve_{s\vk}}=f^i_{s\vk}+\phi^i_{s\vk}\frac{f^i_{s\vk}(1-f^i_{s\vk})}{T}
\label{eqn:relax}
\enq
where $\phi^i_{s\vk}=e\tau^i_\vk\EE_i\cdot \vv_{s\vk}$,
$\tau^i_\vk=\tau^i_k$ the transport
scattering time in the layer $i$,
$f^i_{s\vk}=1/[e^{(\ve_{s\vk}-\mu_i)/T}+1]$ the equilibrium Fermi
distribution function in layer $i$, $\mu_i$ the chemical potential in
the layer $i$, $T$ the temperature of the system. We have set
$\hbar=1$ and the
Boltzmann constant $k_B=1$ throughout this paper. The current in layer $i$ can then be
written as
\beq
\jj_i=-ge\sum_{s\vk}\vv_{s\vk}\tilde{f}^i_{s\vk}=ge\sum_{s\vk}\tau^i_\vk\vv_{s\vk}{\cal I}^i_{s\vk}
\label{eqn:current}
\enq
where $g=4$ is the spin and valley degeneracy of the excitations.  We
will now focus on the form of the collision integral in the presence
of the two layers. The collision integral in each layer has two
contributions: (i) from impurity scattering in the same layer and (ii)
from Coulomb scattering with electrons of the other layer; i.e. $
\Ical^i_{s\vk}=\Ical^{i(imp)}_{s\vk}+\Ical^{i(C)}_{s\vk}$ and hence
$\jj_i=\jj_i^{(imp)}+\jj_i^{(C)}$. The impurity scattering
contribution to the current $\jj_i^{(imp)}=\sigma^i\EE_i$, $\sigma^i$
being the usual conductivity of the layer $i$ in absence of the second
layer. From now on, we will focus on the Coulomb
scattering contribution and drop the superscript $(C)$ in the
collision integral and the current.

The electron-electron scattering between the layers is mediated
through a dynamically screened inter-layer Coulomb interaction
$V(\qq,\omega)$. We will discuss the precise form of the screened
Coulomb interaction at the end of this section. The collision integral
is given by
\begin{widetext}
\bqa
\nonumber \displaystyle \Ical^i_{s\vk}&=&{2\pi g}\sum_{s'r,r'}\sum_{\kp\qq}\int d\omega |V(q,\omega)|^2F^{ss'}_{\vk,\qq}F^{rr'}_{\kp,-\qq}\left[(1-\tilde{f}^i_{s\vk})\tilde{f}^i_{s'\vk+\qq}(1-\tilde{f}^l_{r\kp})\tilde{f}^l_{r'\kp-\qq}
- \tilde{f}^i_{s\vk}(1-\tilde{f}^i_{s'\vk+\qq})\tilde{f}^l_{r\kp}(1-\tilde{f}^l_{r'\kp-\qq})\right]\\
 & &~~~~~~~~~~~~~~~~~~~~~~~~~~~~~~\times \delta(\omega+\ve_{s\vk}-\ve_{s'\vk+\qq})\delta(\omega+\ve_{r'\kp-\qq}-\ve_{r\kp})
\label{eqn:coll:1}
\eqa
where $r,r'=\pm 1$, $l$ is the layer other than $i$, and $F^{ss'}_{\vk,\qq}=|\psi^\dagger_{s'\vk+\qq}\psi_{s\vk}|^2$ is the wavefunction overlap between the bands.
\end{widetext}

It is easy to verify that the collision integral vanishes if we
replace $\tilde{f}^i$ by the equilibrium approximation, $\tilde{f}^i=f^i$. Using
Eq.~(\ref{eqn:relax}), to linear order in the deviations from the
equilibrium distribution, the collision integral is given by
\begin{widetext}
\bqa
\nonumber \Ical^i_{s\vk}&=&\frac{2\pi g}{T}\sum_{s'r,r'}\sum_{\kp\qq}\int d\omega |V(q,\omega)|^2F^{ss'}_{\vk,\qq}F^{rr'}_{\kp,-\qq}f^i_{s\vk}(1-f^i_{s'\vk+\qq})f^l_{r\kp}(1-f^l_{r'\kp-\qq})(\phi^i_{s\vk}-\phi^i_{s'\vk+\qq}+\phi^l_{r\kp}-\phi^l_{r'\kp-\qq})\\
& &~~~~~~~~~~~~~~~~~~~~~~~~~~~~~~\times \delta(\omega+\ve_{s\vk}-\ve_{s'\vk+\qq})\delta(\omega+\ve_{r'\kp-\qq}-\ve_{r\kp})
\label{eqn:coll:2}
\eqa
\end{widetext}
Multiplying Eq.~(\ref{eqn:coll:2}) by $eg\tau^i_\vk\vv_{s\vk}$ and
summing over $s$ and $\vk$, the terms with $\phi^i$ vanishes. Further,
using the identity
\beq
f_{s\vk}(1-f_{s'\kp})\delta(\omega+\ve_{s\vk}-\ve_{s'\kp})=\frac{f_{s\vk}-f_{s'\kp}}{1-e^{-\frac{\omega}{T}}}
\enq
we
get the current in layer $i$,
\begin{widetext}
\bqa
\jj_i&=&\frac{\pi g^2e^2}{2 T}\sum_{\qq}\int d\omega \frac{|V(\qq,\omega)|^2}{\sinh^2\frac{\omega}{2T}} \sum_{ss'\vk}F^{ss'}_{\vk,\qq}\tau^i_\vk\vv_{s\vk}(f^i_{s\vk}-f^i_{s'\kp})\delta(\omega+\ve_{s\vk}-\ve_{s'\vk+\qq})\\
\nonumber & &\times \sum_{rr'\kp}F^{rr'}_{\kp,\qq}\EE_l\cdot(\tau^l_\kp\vv_{r\kp}-\tau^l_{\kp+\qq}\vv_{r'\kp+\qq})(f^l_{r\kp}-f^l_{r'\kp+\qq})\delta(\omega+\ve_{r\kp}-\ve_{r'\kp+\qq})
\eqa
\end{widetext}
Rearranging the terms with $f^i$, we finally obtain
\beq
\jj_i^\alpha=\sigma_D\EE_l^\alpha,
\label{eqn:drag:rho}
\enq
where
\beq
\sigma_D=\frac{1}{4\pi T}\sum_\qq\int d\omega \frac{|V(\qq,\omega)|^2 Im[\chi^\alpha_a(\qq,\omega)] Im[\chi^\alpha_p(\qq,\omega)]}{\sinh^2\frac{\omega}{2T}}
\label{eqn:drag:sigma}
\enq
and the non-linear drag susceptibility is given by
\beq
\chi_i(\qq,\omega)=e\sum_{ss'\vk}\frac{F^{ss'}_{\vk,\qq}(f^i_{s\vk}-f^i_{s'\vk+\qq})(\tau^i_\vk\vv_{s\vk}-\tau^i_{\vk+\qq}\vv_{s'\vk+\qq})}{\omega+\ve_{s\vk}-\ve_{s'\vk+\qq}+i0^+}.
\label{eqn:nonlinsusc}
\enq
Then the drag resistivity is given by
\begin{equation}
\rho_D=-\frac{\sigma_D}{\sigma^a\sigma^p-\sigma_D^2}\sim
  -\frac{\sigma_D}{\sigma^a\sigma^p}.
\end{equation}

\begin{figure}[t]
 \includegraphics[width=\columnwidth]{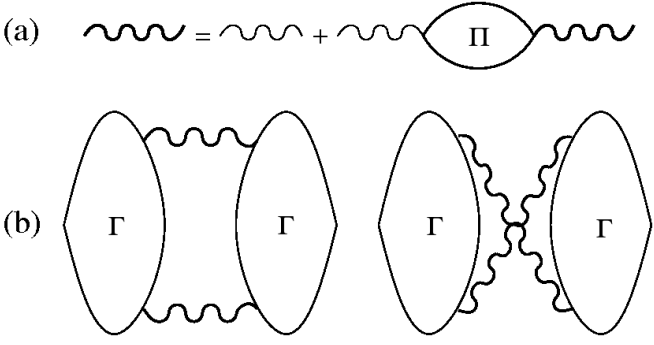}
 \caption{
(a) Screened interlayer Coulomb interaction in the RPA. The thin and thick lines are
the bare and the screened interactions, respectively. The bare bubble
represents the polarizability $\Pi(q,\omega)$.
(b) The leading order diagrams contributing to the drag
conductivity. $\Gamma$ indicates the non-linear susceptibility.
 \label{diagram}
 }
 \end{figure}

We will now focus on the last piece of information required to
calculate the drag resistivity of chiral electron-hole systems, the
dynamically screened interaction, $V(\qq,\omega)$, as shown in
Fig.~\ref{diagram}(a). The intra-layer 
bare Coulomb interaction is given by $V_{aa}(q)=V_{pp}(q)=2\pi
e^2/\kappa q$, where $\kappa$ is the dielectric constant. The
inter-layer bare Coulomb interaction is given by
$V_{ap}(q)=V_{pa}(q)=V_{aa}(q)e^{-qd}$, where $d$ is the layer
separation. Within random phase approximation (RPA), the dynamically
screened inter-layer interaction is given by
$V(\qq,\omega)=V_{ap}(q)/\epsilon(\qq,\omega)$, with the dielectric function of coupled layer systems given by~\cite{hwang:2009} 
\begin{eqnarray}
|\epsilon(q,\omega)| & = & \left [1-V_{aa}(q)\Pi_a(q,\omega) \right ]
\left [1-V_{pp}(q)\Pi_p(q,\omega) \right ] \nonumber \\
& - & V_{ap}(q)V_{pa}(q)\Pi_a(q,\omega)\Pi_p(q,\omega),
\label{eq:eps}
\end{eqnarray}
where $\Pi_i$ is the polarizability of the layer $i$.
\beq
\Pi_i(\qq,\omega)=-g\sum_{ss'\vk}\frac{F^{ss'}_{\vk,\qq}(f^i_{s\vk}-f^i_{s'\vk+\qq})}{\omega +\ve_{s\vk}-\ve_{s'\vk+\qq}}.
\label{eq:pi}
\enq
Eq.~(\ref{eqn:drag:rho}) - Eq.~(\ref{eq:pi}) thus completely defines
the drag resistivity of chiral electron-hole systems in terms of the
band dispersion and wavefunctions. In the next two sections, we will
adapt these general equations to study the drag resistivity in
monolayer and bilayer graphene.


So far we have used the linearized Boltzmann equation to derive the
drag conductivity which relates the induced electric field in one
layer to the driving current in the other layer in a double layer
system. Before moving on to the specific case of SLG and BLG, we would
like to show that the linearized Boltzmann result captures the leading
order result for drag conductivity within a diagrammatic expansion of
the linear response Kubo formula. 
When an external field is applied to layer 1
and the induced current is measured in layer 2, the drag conductivity
is given by the Kubo formula \cite{flensberg:1995a}
\begin{equation}
\sigma_D(\omega) = \frac{1}{\omega A} \int_0^{\infty}dt e^{i\omega t}
\left \langle \left [J_1^{\dagger}(t),J_2(0) \right ] \right \rangle,
\label{eq:kubo}
\end{equation}
where $A$ is the area of the sample and $J_i$ is the current operator
in the $i$-th layer.

The nonvanishing leading order diagrams corresponding to
Eq.~(\ref{eq:kubo}) are given in Fig.~\ref{diagram}(b). The two leading
order diagrams can be written in a symmetric form
\begin{eqnarray}
\sigma_D(i\Omega_n) = \frac{1}{2i\Omega_n}\frac{1}{T}
\sum_{q,i\omega_m} \Gamma_1(q,\omega_m+\Omega_n)\Gamma_2(q,\omega_m)
\nonumber \\
\times V(q,\omega_m+\Omega_n)V(q,\omega_m),
\end{eqnarray}
where $\Omega_n=2\pi inT$ and $\omega_m=2\pi imT$ are boson
frequencies, $V(q,\omega)$ is the interlayer screened 
Coulomb interaction [Fig.~\ref{diagram}(a)], and 
$\Gamma_i(q,\omega)$ is the three-point vertex diagrams (or the
non-linear susceptibility) and given by \cite{flensberg:1995a,kamenev:1995,tse:2007a}
\begin{eqnarray}
\Gamma(q,\omega) = T \sum_{\epsilon_n}{\rm
  Tr} \left \{ G_{\epsilon_n}G_{\epsilon_n + \omega} J(q)
G_{\epsilon_n+\omega} \right \} \nonumber \\
+  {\rm Tr} \left \{ G_{\epsilon_n}G_{\epsilon_n - \omega} J(q)
G_{\epsilon_n - \omega} \right \},
\end{eqnarray}
where $G_{\epsilon_n}$ is the Green function, $J(q)$ stands for the
current and ``Tr'' the trace.  To get the dc
drag conductivity we need to perform an analytical continuation of
external frequencies to a real value, $i\Omega_n \rightarrow \Omega$, and the limit
$\Omega \rightarrow 0$ should be taken.

After summing over the boson frequencies, $\omega_m$, and performing
an analytical continuation to a real value of $\Omega$ we have
\begin{equation}
\sigma_D = \frac{1}{16\pi T}\sum_q\int
\frac{dw}{\sinh^2\frac{\omega}{2T}} \Gamma_1(q,\omega)
\Gamma_2(q,\omega) \left | V(q,\omega) \right |^2.
\end{equation}
The nonlinear susceptibility with real frequencies is given by
\begin{eqnarray}
\Gamma(q,\omega) & = & \frac{1}{4\pi i}\int d\epsilon  \left (
\tanh\frac{\epsilon}{2T} - \tanh \frac{\epsilon}{2T} \right ) \nonumber
\\
& \times & \sum_{p}{\rm Tr} \left [ \left ( G_{\epsilon}^{-}-G_{\epsilon}^{+} \right )
  G_{\epsilon+\omega}^{-}J(p) G_{\epsilon+\omega}^{+} \right ]
\nonumber \\
& + & \left \{ (q,\omega) \rightarrow (-q,-\omega) \right \},
\end{eqnarray}
where $G_{\epsilon}^{\pm} = (\epsilon-H \pm i\gamma)^{-1}$ denotes the
retarded ($-$) and advanced ($+$) Green function for a given system
with the Hamiltonian $H$, respectively. Here we use a damping
constant $\gamma = 1/2\tau$ to include the disorder scattering in the Green function.
In the Boltzmann regime ($\omega \tau \gg 1$ or $k_F l \gg 1$, where
$k_F$ is the Fermi wave vector and 
$l=v_F \tau$ is the mean free path), which corresponds to weak
impurity scattering and the actual experimental regime of
the high mobility graphene samples, we can treat the vertex correction
in the current $J(q)$ within
the impurity ladder approximation \cite{kamenev:1995,tse:2007a}. 
Then, the impurity-dressed current vertex becomes $J = (\tau_{tr}/\tau)
\partial H/\partial k $, where $\tau_{tr}$ is the transport time. 
By using the following equation 
\begin{equation}
G_{\epsilon+\omega}^-(p)G_{\epsilon+\omega}^+(p) = 2 \tau {\rm Im}
G_{\epsilon +\omega}^+(p),
\end{equation}
and expressing the matrix form of the Green function in the chiral
basis, finally, we have the nonlinear susceptibility 
\begin{eqnarray}
\Gamma(q,\omega) = \tau \sum_{ss', k} \left [ J_{s}(k) - J_{s'}(k+q)
  \right ] F_{ss'} (k,k+q) \nonumber \\
\times {\rm Im} \frac{f_{sk}-f_{s'k+q}}{\omega +
  \epsilon_{sk}-\epsilon_{s'k+q}+i0^+}.
\end{eqnarray}
Comparing the above equation with Eq.~(\ref{eqn:nonlinsusc}) we have
\begin{equation}
\Gamma(q,\omega) = 2{\rm Im} \chi (q,\omega).
\end{equation}
Thus we recover the Boltzmann equation drag conductivity result by
employing the leading order diagrammatic expansion within the Kubo
formalism.


\section{Drag Resistivity in Monolayer Graphene}

Monolayer graphene (MLG) is characterized by the presence of gapless
linearly dispersing electron and hole bands with dispersion,
\beq
\epsilon_{s\vk}=v_Fk,
\enq
where $s=\pm 1$ denotes the electron and hole bands and $v_F$ is the
Fermi velocity of graphene. The linear dispersion of MLG implies that
the group velocity of the bands are given by
$v_{s\vk}=v_{F}\vk/|\vk|$. Thus, contrary to usual semiconductor
systems, or the case of bilayer graphene to be treated later, the
group velocity is constant in magnitude and does not scale with the
momentum of the band-states. As we will show, this leads to profound
qualitative differences in the variation of the MLG drag resistivity
with the carrier densities in the layers and the distance between the
layers. The chirality of the electrons (holes) are encoded in the band
wavefunctions
\beq
\psi_{s\vk}=\frac{1}{\sqrt{2}}\left(\begin{array}{c} e^{-i\theta_\vk}\\s \end{array}\right)
\enq
where $\theta_\vk$ is the azimuthal angle in the 2D $\vk$ space. This
leads to the wavefunction overlap factor
$F^{ss'}_{\vk,\qq}=(1/2)[1+ss'\cos (\theta_{\vk+\qq}-\theta_{\vk})]$.
 
Before we go on to a detailed discussion of the non-linear drag
susceptibility and the drag resistivity in MLG, let us first discuss
the finite temperature polarizability in MLG which will control the
finite temperature screening of the Coulomb potential.
Although analytic expressions for graphene polarizability at $T=0$ has
been worked out before\cite{hwang:2007}, the finite temperature
versions of the polarizability have not been calculated
analytically. The full expression of finite temperature polarizability
is necessary to understand more precisely the temperature dependent
drag including the plasmon enhancement effects\cite{flensberg:1995}.
Here we provide a finite temperature generalization of our earlier
work on zero temperature graphene
polarizability~\cite{hwang:2007}. The MLG polarizability of layer $i$
is given by
\begin{widetext}
  \bqa
  \Pi_i(q,\omega,) &=& \frac{2E_{Fi}}{\pi v_F^2}\left(\frac{\pi}{8}\frac{y^2}{\sqrt{y^2-z^2}}\right.\\
  \nonumber &+&\left.\int_{0}^{\infty}dx \left[ f(x) +g(x) \right ] \left [
    \frac{ \left ( \frac{z^2-y^2}{4}+zx + x^2 \right ) {\rm
        sgn}(a_{+}) }{ \sqrt{\frac{(z^2-y^2)^2}{4} + (z^2-y^2) (z x +
        x^2)}} + \frac{\left ( \frac{z^2-y^2}{4}-z x+ x^2 \right )
      {\rm sgn}(a_{-}) }{ \sqrt{\frac{(z^2-y^2)^2}{4} + (z^2-y^2) (-z
        x + x^2)}} \right ]\right),
\label{eq:pit}
\eqa 
where $x=k/k_{Fi}$, $y=q/k_{Fi}$ and $z=\omega/E_{Fi}$, $f(x) =
[e^{-(x-\mu_i)/t_i}-1]^{-1}$, $g(x) = [e^{(x+\mu_i)/t_i}+1]^{-1}$,
$a_{\pm} = z^2 - y^2 \pm 2zx$, with $E_{Fi}$ and $k_{Fi}$ being the
Fermi energy and the Fermi wave-vector in layer $i$. Here, $\mu_{i}$
is the chemical potential in layer $i$ in units of $E_{Fi}$ (to be
calculated self-consistently) and $t_i=T/E_{Fi}$.
\end{widetext}


We now turn our attention to the non-linear drag susceptibility and
drag resistivity in MLG systems. The drag resistivity in MLG crucially
depends on the variation of the transport scattering time with energy
(or equivalently momentum). In this paper, we will consider the drag
resistivity of MLG using two different models of scattering time: (a)
a momentum independent scattering time $\tau_{\vk}=\tau$ and (b) a
scattering time scaling linearly with momentum (energy),
$\tau_\vk=\tau_0|\vk|$, which results from the unusual screening of
charge impurity potential in this linearly dispersive
material~\cite{novoselov:2005,tan:2007,chen:2008,dassarma:2011}.



We first consider the energy independent scattering time approximation. In this case, the non-linear drag susceptibility in MLG can be written as
\begin{widetext}
\beq
\chi_i(q,\omega)=4\tau_i
v_F\sum_{ss'\vk}\left(s\frac{\vk+\qq}{|\vk+\qq|}-s'\frac{\vk}{|\vk|}\right)
\frac{F^{ss'}_{\vk,\qq}(f^i_{s\vk+\qq}-f^i_{s'\vk})}{\omega-\ve_{s\vk+\qq}+\ve_{s'\vk}+i0^+}
\enq 
\end{widetext}
Within the energy independent scattering time approximations, the
intra-layer conductivities are given by
$\sigma_i=e^2E_{Fi}\tau_i/4$.

We focus on the analytic asymptotic behaviors of the drag resistivity
in MLG at low temperatures, both in the large layer separation weak
coupling limit ($k_Fd \gg 1$) and in the small layer separation strong
coupling limit($k_Fd \ll 1$).  To calculate the asymptotic behavior of
drag resistivity first we investigate the non-linear susceptibility
for $\omega <  v_F q < E_F$. Due to the phase-space restriction
the most dominant contribution to the drag resistivity arises from
$\omega <  v_F q$ at low temperatures.  When we neglect the
energy dependence in the transport times, i.e., $\tau_k=\tau$, then
we obtain, for $\omega <  v_F q$,
\begin{equation}
\chi(q,\omega) \sim
\frac{\tau q^2}{\pi E_F}\frac{\omega}{v_Fq}\sqrt{1-{q^2}/{4k_F^2}}.
\end{equation}
With assumptions of a large
inter-layer separation ($k_{F}d\gg 1$, or $q_{TF}d\gg 1$, with $q_{TF}$
being the Thomas Fermi (TF) screening wave 
vector) and the random phase approximation (RPA) in which $\Pi
_{ii}$ is replaced by its value for the non-interacting electrons, we have 
the drag resistivity at high density and low temperature,
\begin{equation}
\rho_D = \frac{h}{e^2}\frac{5! \zeta(5)}{3\cdot2^8}
\frac{(k_BT)^2}{E_{F_1} E_{F_2}}\frac{1}{(q_{TF_1}d) (q_{TF_2}d)
  (k_{F_1}d)^2 (k_{F_2}d)^2}
\label{eq:anal1}
\end{equation}
where $q_{TF}=4 r_s k_F$ 
is the TF wave vector with the graphene fine structure constant
$r_s = e^2/\kappa  v_F$ 
and $\zeta(x)$ is the Riemann zeta function. 
This result shows that $\rho_D(n) \propto n^{-4}$ and $\rho_D(T)
\propto T^{2}$. 
For large layer separation (i.e. $k_Fd \gg 1$)
the back-scattering $q \approx 2k_F$ is suppressed
due to the exponential dependence of the interlayer Coulomb
interaction $v_{12}(q) \propto \exp(-qd)/q$ as well as the graphene
chiral property. In this case the drag is
dominated by small angle scattering and one expects $\rho_D \propto
T^2/(n^4 d^6)$. 

For the strong interlayer correlation ($k_F d \ll1$) in the
low-density or small-separation limit, the asymptotic behavior
behavior of drag resistivity becomes
\begin{equation}
\rho_D = \frac{h}{e^2}\frac{1}{6}\frac{(k_BT)^2}{E_{F_1}E_{F_2}}
\frac{r_s^2}{(k_{F_1}d)(k_{F_2}d)}.
\end{equation}
We have the same temperature dependence, $\rho_D(T) \sim T^2$,
but the density dependence becomes much weaker, $\rho_D(n) \sim
1/(nd)^2$. At low
densities (or strong interlayer correlation, $k_F d \ll 1$) the exponent
in the density dependent drag  differs from -4.
In an ordinary 2D systems,
at low carrier densities and for closely spaced layers
the backward scattering can be important since $k_Fd \sim 1$.
In the low temperature range $T/T_F \ll 1$ the enhanced phase space
for $q=2k_F$ backward Coulomb scattering leads to $\ln(T)$ corrections
to the usual $T^2$ dependence of the drag, i.e. $\rho_D \propto T^2\ln T$. 
However, due to the suppression of the $q=2k_F$
back-scattering due to the chirality of graphene there is no
$\ln(T)$ correction in the drag 
resistivity of monolayer graphene. 


 \begin{figure}
 \includegraphics[width=\columnwidth]{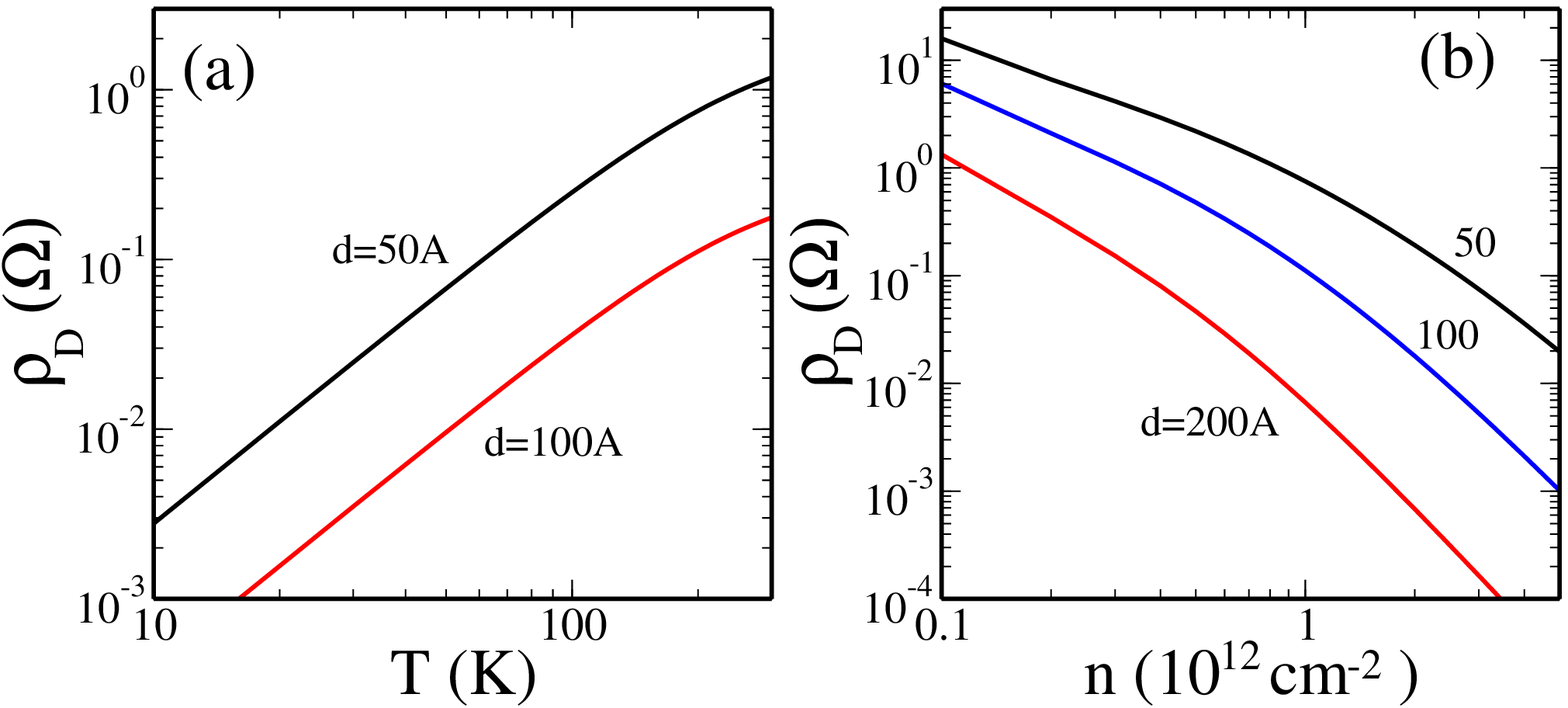}
 \caption{The calculated drag resistivity by considering the energy
   independent scattering approximation.
(a)  The temperature dependence of Coulomb drag for two different layer separations
  $d=50$\AA, $d=100$\AA\;  and the equal
 electron densities, $n_1=n_2=10^{12}cm^{-2}$. 
(b)  The density dependent Coulomb drag 
 for different layer separations $d=5$, 10, 20 nm and at $T=200K$.
 \label{fig:1}
 }
 \end{figure}

The drag resistivity within this
approximation is plotted as a function of temperature and 
density in Fig.~\ref{fig:1} for different layer separations. The
parameters corresponding to the experimental setup of
ref.~\onlinecite{kim:2011} are used.
In Fig.~\ref{fig:1}(a) we show the calculated Coulomb drag as a function
of temperature for an equal carrier density,   
$n_1=n_2= 10^{12}cm^{-2}$, for two different layer separations $d=50$ {\AA} and
$d=200$ {\AA}. The overall temperature dependence of drag is  
close to the quadratic behavior, $\rho_D \propto T^2$. But we find a small
corrections at low temperatures, especially at low values of $k_Fd$.
In regular 2D systems there is  a $\ln(T)$ corrections to the $T^2$ dependence
of the drag. However, due to the suppression of the back-scattering in
graphene such logarithmic
correction does not show up in our numerical results except perhaps at
extremely low temperatures. 
In Fig.~\ref{fig:1} (b) the density dependent Coulomb drag is shown for
different layer separations.
Our calculated Coulomb drag resistivity 
follow a $n^{\alpha}$ dependence with $\alpha \sim -2$ at low carrier
densities (or, $k_F d <1$), 
but as the density increases the exponent decrease to $\alpha\sim-4$.


So far, we have considered the energy independent scattering
time. However, it is known that the scattering by the charged impurity
disorder which inevitably exist in the graphene environment dominates
and the scattering time due to the charged impurity is linearly
proportional to the energy, $\tau_i \sim \ve=\tau^0_i
k$. \cite{novoselov:2005,tan:2007,chen:2008,dassarma:2011}. In this
approximation, the nonlinear drag susceptibility is given by
\begin{widetext}
\beq
\chi_i(q,\omega)=4\tau^0_i v_F\sum_{ss'\vk}[s(\vk+\qq)-s'\vk]\frac{F^{ss'}_{\vk,\qq}(f^i_{s\vk+\qq}-f^i_{s'\vk})}{\omega-\ve_{s\vk+\qq}+\ve_{s'\vk}+i0^+}
\enq
\end{widetext}
In this case, we will mainly focus on the low temperature asymptotic for of the drag resistivity. For linearly energy 
dependent scattering time due to the charged impurities,
the non-linear susceptibility for $w< v_F q$ is
\begin{equation}
\chi_i(q,\omega) = \frac{2\tau^0_i}{\pi}\frac{k_F}{ v_F}\frac{\omega}{E_F}
\frac{1}{\sqrt{1-q^2/4k_F^2} }.
\end{equation}
Then the drag resistivity for a large inter-layer separation ($k_Fd
\gg 1$) is given by
\begin{equation}
\rho_{D}=\frac{h}{e^2}
\frac{\zeta(3) }{2^3}
  \frac{(k_BT)^2}{E_{F_1}E_{F_2}}\frac{1}{(k_{F_1}d)(k_{F_2}d)(q_{TF_1}d)
  (q_{TF_2}d)} ,
\label{eq:anal2}
\end{equation}
and for the strong interlayer correlation limit ($k_{F}d \ll 1$), we
have
\begin{equation}
\rho_D =\frac{h}{e^2}\frac{2^4
  r_s^2}{3}\frac{(k_BT)^2}{E_{F_1}{E_{F_2}}}  \ln \left [
  \frac{2(q_{TF_1}+q_{TF_2})d +1}{2(q_{TF_1}+q_{TF_2})d} \right ].
\end{equation}
Thus we see that when we consider the energy dependent intralayer
scattering time we have very different asymptotic behaviors compared
to the results from energy independent scattering time, i.e., we
find for $k_Fd \gg 1$, $\rho_D \sim T^2/(n^3 d^4)$ and for $k_F d \ll
1$, $\rho_D \sim T^2 \ln (\sqrt{n} d)$. Thus in the presence of the
strong charged impurity scattering the Coulomb drag resistivity follow
a $n^{\alpha}$ dependence with $\alpha \alt -3$ at high densities but
as the density decreases the exponent ($\alpha$) increase.  Based on
our calculation we believe that the experimental departure from the
$n^{-3}$ behavior reported in Ref.~\cite{kim:2011} is essentially a
manifestation of the fact that the asymptotic $n^{-3}$ regime is hard
to reach in low density electron systems where $k_F d \gg 1$ limit
simply cannot be accessed. We predict a weak $\ln(n)/n$ density
dependence in the low-density or small separation limit.

 \begin{figure}
 \includegraphics[width=\columnwidth]{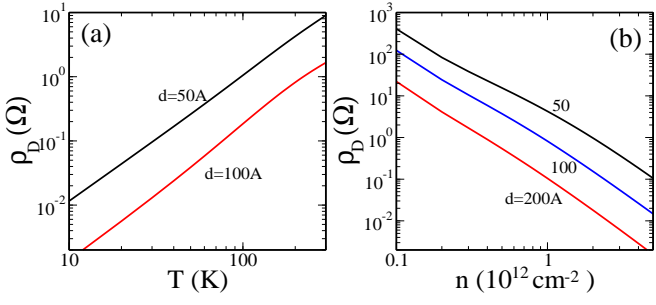}
 \caption{The calculated drag resistivity by considering the linearly energy
   dependent scattering time.
(a)  The temperature dependence of Coulomb drag for two different layer separations
  $d=50$\AA, $d=200$\AA\; and the equal
 electron densities, $n_1=n_2=10^{12}cm^{-2}$. 
(b)  The density dependent Coulomb drag 
 for different layer separations $d=5$, 10, 20 nm and at $T=200K$.
 \label{fig:2}
 }
 \end{figure}

In Fig.~\ref{fig:2}(a) we show the calculated Coulomb drag as a function
of temperature for two different layer separation (a) $d=50$ {\AA} (b)
$d=100$ {\AA} by considering the linearly energy dependent scattering
time. The overall temperature dependence of drag increases
quadratically and there is no logarithmic correction 
due to the suppression of the back-scattering.
In Fig.~\ref{fig:2}(b) the density dependent Coulomb drag is shown for
different layer separations. The density dependent Coulomb drag 
follow a $n^{\alpha}$ dependence with $\alpha \sim -2$.
Based on our calculation the consideration of the energy dependent
scattering time is crucial to understand experiment measurements of
Ref.~\onlinecite{kim:2011}.

 \section{Drag Resistivity in Bilayer Graphene}

 In this section we will study the drag resistance in a
 heterostructure made of two bilayer graphene (BLG) layers separated
 by an insulating barrier and study its dependence on temperature,
 density and separation of the layers. Both BLG and MLG have chiral
 gapless electron and hole bands. However, BLG differs from MLG in two
 crucial aspects: (i) the bands have a parabolic dispersion as opposed
 to linear dispersion in MLG and (ii) the chiral angle is double that
 in MLG leading to enhanced rather than suppressed backscattering
 between the quasiparticles. We will show in this section how these
 two differences lead to dramatic changes in the density and layer
 separation dependence of the BLG drag resistivity compared to MLG
 drag resistivity.

 BLG consists of an electron and a hole band with quadratic dispersion
\beq
\epsilon_{s\vk}=sk^2/2m
\enq
($s=\pm 1$ for electron(hole) band) and the BLG mass is $m=0.033m_e$,
where $m_e$ is the electron mass. The group velocity in the two bands
are $\vv_{s\vk}=s\vk/m$, which scales linearly with momentum. The 
wavefunctions in these bands are given by
\beq
\psi_{s\vk}=\frac{1}{\sqrt{2}}\left(\begin{array}{c} e^{-2i\theta_\vk}\\s \end{array}\right)
\enq
which gives the overlap factor $F^{ss'}_{\vk,\qq}=(1/2)(1+ss')
-ss'q^2\sin^2\phi/|\vk+\qq|^2$, where $\phi$ is the angle between
$\vk$ and $\qq$.

 Without loss of generality, we will assume that the applied
electric field, the induced electric field and the resultant currents
are all in the $\hat{x}$ direction. We will also assume that the
chemical potential in both layers is in the conduction (electron)
band, i.e. they are electron doped. The chemical potential as a function of temperature is obtained by solving the integral equation
\beq
g\sum_{\vk} \frac{1}{e^{(k^2/2m-\mu_i)/T}+1}+\frac{1}{e^{-(k^2/2m+\mu_i)/T}+1}-1=\rho_i
\enq
where $\rho_i$ is the density of the layer $i$. It is an interesting
feature of the BLG dispersion that in the non-interacting
approximation, the chemical potential is independent of the
temperature and is given by $\mu_i=E_{fi}$, where $E_{fi}$ is the
Fermi energy of the electrons in layer $i$.

We will first focus on the polarizability and hence the screened
Coulomb interaction in BLG systems. The analytic form for the BLG
polarizability at zero temperature has been derived
before~\cite{sensarma:2010}, but we need to take into account the
temperature dependence of the screening to study the detailed
behaviour of the drag resistivity with temperature. Here, we
generalize our earlier results on BLG polarizability to finite
temperature. The polarizability can be broken up into the intra-band
[$s=s'$ terms in Eq.~(\ref{eq:pi})] and inter-band [$s\neq s'$ terms in
Eq.~(\ref{eq:pi})] contributions. Working out the azimuthal integrals
analytically we obtain 
\begin{widetext}
\beq
\Pi_i^{intra}(\qq,\omega)=\frac{gm}{4\pi}\int_0^\infty dx\frac{(f^i_{+x}-f^i_{-x})}{x(x^2+z)}\left[|x^2-y^2|-(x^2+z)+sgn(\zeta_1+2xy)\frac{(2x^2+\zeta_1)^2}{\sqrt{\zeta_1^2-4x^2y^2}}\right]+(y,z \rightarrow -y,-z)
\enq
and
\beq
\Pi_i^{inter}(\qq,\omega)=-\frac{gm}{4\pi}\int_0^\infty dx\frac{(f^i_{+x}-f^i_{-x})}{x(x^2+z)}\left[|x^2-y^2|+(x^2+z)-sgn(\zeta_2+2xy)\sqrt{\zeta_2^2-4x^2y^2}\right]+(y,z \rightarrow -y,-z)
\enq
where $y=q/k_{Fi}$, $x=k/k_{Fi}$, $z=\omega/E_{Fi}$, $\zeta_1=z-y^2$, $\zeta_2=z+2x^2+y^2$, and $f_{sx}=1/(e^{(sx^2-1)/t_i}+1)$ with $t_i=T/E_{Fi}$
\end{widetext}

We will now shift our attention to the non-linear drag
susceptibility. For BLG systems, both charge impurity scattering and
short range impurity scattering leads to a transport scattering time
independent of momenta, and hence, we will only consider an energy
independent scattering time, $\tau^i_{\vk}=\tau^i$, in this case. With this approximation the nonlinear drag susceptibility is given by
\begin{widetext}
 \beq
  \chi_i(q,\omega)=\frac{4\tau_i}{m}\sum_{ss'\vk}[s(\vk+\qq)-s'\vk]\frac{F^{ss'}_{\vk,\qq}(f^i_{s\vk+\qq}-f^i_{s'\vk})}{\omega-\ve_{s\vk+\qq}+\ve_{s'\vk}+i0^+}
  \enq 
  where $m$ is the BLG mass. $m=0.033m_e$, where $m_e$ is the free
  electron mass.
\end{widetext}

The
non-linear susceptibility $\chi^{x}_i$ can be separated into an
intra-band contribution and an inter-band contribution. The azimuthal
integrals can be done analytically and we get
\begin{widetext}
\beq
Im[\chi^{x(intra)}_i(q,\omega)]=\frac{\tau^ik_{Fi}y\cos \phi_q}{\pi}\left[\int_{\frac{|z-y^2|}{2y}}^\infty dx\frac{f^i_{+x}+f^i_{-x}}{x(x^2+z)}\frac{(2x^2+z-y^2)^2}{\sqrt{4x^2y^2-(z-y^2)^2}} - (z\rightarrow -z) \right]
\enq
and 
\bqa
\displaystyle Im[\chi^{x(inter)}_i(q,\omega)]=\frac{\tau^ik_{Fi}\cos \phi_q}{\pi y}\left[\Theta(b)\int_{\frac{|y-\sqrt{b}|}{2}}^{\frac{|y+\sqrt{b}|}{2}} dx\frac{f^i_{+x}+f^i_{-x}}{x(x^2-z)}(2x^2-z)\sqrt{4x^2y^2-(2x^2+y^2-z)^2}-(z\rightarrow -z)\right]
\eqa
where $b=2z-y^2$ and $\phi_q$ is the azimuthal angle related to the
vector $\qq$.
\end{widetext}
\begin{figure*}
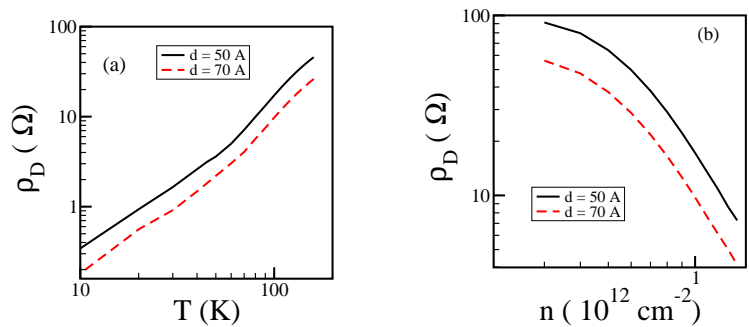

\includegraphics[width=0.24\textwidth]{final_fig4a.eps}~~~~~~~~~~
\includegraphics[width=0.24\textwidth]{final_fig4b.eps}
\caption{(a): The Coulomb drag resistivity in BLG as a function of temperature for two layer separations, $d=50 \AA$ (thick black line) and $d=70 \AA$ (dashed red line). The density of the two layers $n_1=n_2=10^{12}cm^{-2}$. (b): The density dependence of bilayer Drag at a temperature $T=100 K$. The thick black line is for $d=50 \AA$ and the dashed red line is for $d=70 \AA$. We have used a dielectric constant $\kappa=4$ corresponding to boron nitride substrate}
\label{fig:BLdrag}
\end{figure*}
We note that the intra-band drag susceptibility is proportional to the
intra-band polarizability of BLG
($\chi^{intra}(\qq,\omega)=(4\tau/m)\qq\Pi^{intra}(\qq,\omega)$) only
in the $T=0$ limit. This relation breaks down at finite temperatures.
Finally, with a momentum independent scattering time, and a quadratic
dispersion, the intra-layer conductivity can be written in the simple
form $\sigma^i=n_ie^2\tau^i/m$. Note that the scattering times cancel
in the expression for drag resistivity which is purely dominated by
electron-electron interactions. In figure~\ref{fig:BLdrag} (a), we
plot the temperature dependence of the BLG drag resistivity for two
different layer separations with a common layer carrier density of
$n_1=n_2=10^{12} cm^{-2}$. The drag shows a quadratic behaviour with
logarithmic corrections at low temperatures. In
figure~\ref{fig:BLdrag}(b), we plot the density dependence of the
bilayer drag at $T=100 K$ for two different layer separations.

We now focus on the low temperature
asymptotic behaviour of the BLG drag resistivity.
To obtain the leading temperature dependence of $\rho_D$ at low
temperatures, both $\chi$ and $\Pi$ can be replaced by their intra-band
contributions at $T=0$ in the limit of small $q,\omega$, i.e.
$Im[\chi^i(q,\omega)]\sim (gm/2\pi)\tau_i\omega/k_F$, and the screened
Coulomb interaction is replaced by its static value with
$\Pi_1=\Pi_2\sim gm/(2\pi)$. The screening in BLG is controlled by the
Thomas-Fermi wave-vector $q_{TF}=e^2gm/\kappa$, where $\kappa$ is the
background dielectric constant. For large layer separation,
i.e. $q_{TF}d\gg 1$, the screened interaction has the form $V(q)
\sim q/[D_0q_{TF}\sinh(qd)]$ in which $D_0=gm/2\pi$ is the
density of states of bilayer graphene at Fermi level, and the leading
order drag-resistivity is given by
\beq
\rho_D\sim \frac{1}{e^2}\frac{\pi^2\zeta(3)}{16}
\frac{T^2}{E_{F1}E_{F2}} 
\frac{1}{(k_{F1}d)(k_{F2}d)(q_{TF}d)^2}.
\enq
We note that contrary to MLG, $q_{TF}$ in BLG is a constant and
is independent of the density. Thus the large layer separation limit
is not the same as the high density limit ($k_Fd \gg 1$). In the
opposite limit of small layer separation ($q_{TF}d \ll 1$) and strong
interlayer correlations, the screened interaction takes the form
$V(q) \sim (q_{TF}/D_0)exp(-qd)/(q+2q_{TF})$ and the leading order
drag resistivity is given by 
\beq
\rho_D\sim \frac{1}{e^2}\frac{\pi^2}{24}
\frac{T^2}{E_{F1}E_{F2}} \frac{q_{TF}^2}{k_{F1}k_{F2}} [
  -\ln q_{TF}d +\gamma-1-\ln 4+...],
\enq
where $\gamma$ is the Euler constant. Thus the drag resistance $\sim
T^2$ both in the large and small layer separation limit. The density
dependence of the coefficient of the $T^2$ term is $1/(n_1n_2)^{3/2}$
in both limits. In the large separation limit, the leading order term
$\sim 1/d^4$, whereas in the small separation limit, the drag
resistivity shows a weak logarithmic dependence on the layer
separation. We also note that, since backscattering is enhanced in
BLG, there will be an additional $T^2\ln T$ correction to the formulae
derived here.  

\section{Conclusion}

In this paper, we have theoretically studied the frictional drag between two
spatially separated MLG and BLG layers to lowest nonvanishing order in
the screened interlayer electron-electron interaction using Boltzmann
transport theory (which is equivalent to a leading-order diagrammatic
perturbation theory). We find that the low temperature drag mostly shows
a quadratic temperature dependence, both in MLG and BLG, regardless of
the layer separation and density of carriers. However the density and
layer separation dependence of the coefficient of the $T^2$ term is
very different for MLG and BLG.  

The density and layer separation dependence of low temperature MLG
drag resistivity crucially depends on the variation of the intralayer
momentum scattering time with energy. For energy independent intralayer
scattering time (which does not correspond to any model of disorder in MLG, but captures the effects of both short range and screened Coulomb impurities in standard 2DEG), the drag varies from $\rho_D \propto T^2/(nd)^{2}$
for $k_F d \ll 1$ to $\rho_D \propto T^2/(n^{4}d^{6})$ for $k_F d \gg
1$. However, for energy dependent intralayer scattering times
(corresponding to intralayer Coulomb scattering by random charged
impurities in the environment) the
power law of density dependence is significantly changed. Thus, an
accurate measurement of interlayer MLG drag is in principle capable of
distinguishing the main disorder scattering.
In most
currently available graphene samples the scattering due
to charged impurity disorder dominates. In this case the
intralayer scattering time depends linearly on the energy and the drag
resistivity becomes $\rho_D \propto \ln(\sqrt{n}d)/n$ for $k_F d \ll
1$ and $\rho_D \propto n^{-3}d^{-4}$ for $k_F d \gg 1$.  We note that
the density dependence of drag resistivity is very sensitive to the
experimental setup, so the density dependence does not have any
universal power law behavior
because one is never in any asymptotic regime with clear cut
analytical power law behavior.  Experimental measurements\cite{kim:2011}
therefore may not find any clear cut power law behavior in the density
dependence of MLG drag since one is always in the crossover regime. 
We also find that due to the
suppression of the $q=2k_F$ back-scattering in graphene there are no $\sim
T^2\ln(T)$ corrections to the low temperature MLG drag resistivity.

We would like to take this opportunity to compare this work, which is
a more complete and updated version of ref.~\onlinecite{hwang_sds:drag}, with
recent works on low temperature MLG drag
resistivity\cite{tse:2007,narozhny:2007,peres:2011,katsnelson:2011},
which predict widely varying density and layer separation dependence of
the low temperature MLG drag resistivity. Within energy independent
scattering time approximation, Tse {\it et al.}~\cite{tse:2007}
studied the drag in MLG in the high density limit and obtained
$\rho_D\sim T^2/n^3d^4$. Katsnelson~\cite{katsnelson:2011} also
obtained the same result in the high density limit ($k_Fd \gg 1$) and additionally
considered the $k_Fd \ll 1$ limit of ref.~\onlinecite{tse:2007} finding
a $\rho_D\sim T^2 \ln (nd^2)/n$ dependence in the low density
limit. Narozhny~\cite{narozhny:2007}, on the other hand, found
that the drag vanishes for linearly dispersive Dirac
particles. In the current paper, within the energy independent scattering
time approximation, we find that $\rho_D\sim T^2/n^4d^6$ in the high
density limit and $\rho_D\sim T^2/n^2d^2$ in the low density
limit. Narozhny~\cite{narozhny:2007} computes the drag to
linear order in the inter-layer potential where it is known to vanish
even for ordinary 2D electron systems. The drag resistivity is at
least quadratic in the inter-layer potential and its vanishing to
linear order is a rather trivial result. The main difference between
our present work and Ref.~\onlinecite{tse:2007} and
~\onlinecite{katsnelson:2011} arises because both these papers use a
form of the non-linear drag susceptibility where the band group
velocity scales linearly with the momentum $k$ and results in a
leading order $q$ dependence of the vertex in the
susceptibility. While this is true for quadratic band dispersions as
in regular 2D and BLG systems, the
group velocity in linearly dispersive MLG is constant in
magnitude($v_F$), and combined with the overlap functions, gives rise
to a leading order $q^2$ momentum in the non-linear susceptibility in
our case. This accounts for the discrepancy between our results and
earlier results. 

Finally Peres {\it et. al}~\cite{peres:2011} considers the drag
resistivity of MLG within a linearly energy dependent scattering time
approximation and obtains $\rho_D\sim T^2/n^4d^6$ in the high density
limit and $\rho_D\sim T^2/n^2d^2$ in the low density limit. These
results match with our energy independent scattering time
approximation results, whereas within linearly energy dependent
scattering time, we get $\rho_D\sim T^2/n^3d^4$ in the high density
limit and $\rho_D\sim T^2 \ln (nd^2)/n$ in the low density limit. This
discrepancy can also be understood in terms of the scaling of the
current vertex factors. The vertex factor going into the drag
susceptibility is $v_{s\vk}\tau_\vk$. Peres {\it
  et. al}~\cite{peres:2011} uses a drag susceptibility where 
the vertex factor in the susceptibility
scales as $q^2$.
Thus they recover our energy independent scattering
time scaling. However, with a constant group velocity, as is the case
for MLG, the energy dependent scattering time approximation should
lead to a leading order $q$ dependence (coming from the momentum
dependence of the scattering time) of the vertex and this accounts for
the discrepancy between our results and Peres {\it et. al}.
The conceptual element of our work
is the new role of group velocity in
the nonlinear susceptibility defining the drag, which turns out to be
crucial to the calculation of Coulomb drag in monolayer graphene (with
its linear energy-momentum dispersion).  We have also studied the drag
resistivity in two spatially separated bilayer-graphene structures. We
find that the drag resistivity shows a quadratic temperature
dependence at low temperatures, as for monolayer graphene. The density
dependence of the BLG drag is independent of the layer separation with
$\rho_D \propto T^2/n^{3}$ both in the large and small layer
separation limit. In the large layer separation limit (weak
inter-layer correlation), the drag has a strong $1/d^4$ dependence on
layer separation, whereas this goes to a weak logarithmic dependence
in the strong inter-layer correlation limit.

\section*{Acknowledgment}
The authors gratefully thank Andre Geim for useful discussions and for
asking several penetrating questions. This work is supported by the
US-ONR and NRI-SWAN.


\begin{thebibliography}{10}%
\makeatletter
\providecommand \@ifxundefined [1]{%
 \ifx #1\undefined \expandafter \@firstoftwo
 \else \expandafter \@secondoftwo
\fi
}%
\providecommand \@ifnum [1]{%
 \ifnum #1\expandafter \@firstoftwo
 \else \expandafter \@secondoftwo
\fi
}%
\providecommand \enquote [1]{``#1''}%
\providecommand \bibnamefont  [1]{#1}%
\providecommand \bibfnamefont [1]{#1}%
\providecommand \citenamefont [1]{#1}%
\providecommand\href[0]{\@sanitize\@href}%
\providecommand\@href[1]{\endgroup\@@startlink{#1}\endgroup\@@href}%
\providecommand\@@href[1]{#1\@@endlink}%
\providecommand \@sanitize [0]{\begingroup\catcode`\&12\catcode`\#12\relax}%
\@ifxundefined \pdfoutput {\@firstoftwo}{%
 \@ifnum{\z@=\pdfoutput}{\@firstoftwo}{\@secondoftwo}%
}{%
 \providecommand\@@startlink[1]{\leavevmode}%
 \providecommand\@@endlink[0]{}%
}{%
 \providecommand\@@startlink[1]{%
  \leavevmode
  \pdfstartlink
   attr{/Border[0 0 1 ]/H/I/C[0 1 1]}%
   user{/Subtype/Link/A<</Type/Action/S/URI/URI(#1)>>}%
  \relax
 }%
 \providecommand\@@endlink[0]{\pdfendlink}%
}%
\providecommand \url  [0]{\begingroup\@sanitize \@url }%
\providecommand \@url [1]{\endgroup\@href {#1}{\urlprefix}}%
\providecommand \urlprefix [0]{URL }%
\providecommand \Eprint[0]{\href }%
\@ifxundefined \urlstyle {%
  \providecommand \doi [1]{doi:\discretionary{}{}{}#1}%
}{%
  \providecommand \doi [0]{doi:\discretionary{}{}{}\begingroup
  \urlstyle{rm}\Url }%
}%
\providecommand \doibase [0]{http://dx.doi.org/}%
\providecommand \Doi[1]{\href{\doibase#1}}%
\providecommand \bibAnnote [3]{%
  \BibitemShut{#1}%
  \begin{quotation}\noindent
    \textsc{Key:}\ #2\\\textsc{Annotation:}\ #3%
  \end{quotation}%
}%
\providecommand \bibAnnoteFile [2]{%
  \IfFileExists{#2}{\bibAnnote {#1} {#2} {\input{#2}}}{}%
}%
\providecommand \typeout [0]{\immediate \write \m@ne }%
\providecommand \selectlanguage [0]{\@gobble}%
\providecommand \bibinfo [0]{\@secondoftwo}%
\providecommand \bibfield [0]{\@secondoftwo}%
\providecommand \translation [1]{[#1]}%
\providecommand \BibitemOpen[0]{}%
\providecommand \bibitemStop [0]{}%
\providecommand \bibitemNoStop [0]{.\EOS\space}%
\providecommand \EOS [0]{\spacefactor3000\relax}%
\providecommand \BibitemShut [1]{\csname bibitem#1\endcsname}%
\bibitem{dassarma:2011}%
  \BibitemOpen
  \bibfield{author}{%
  \bibinfo {author} {\bibfnamefont{S.}~\bibnamefont{Das~Sarma}}, \bibinfo
  {author} {\bibfnamefont{S.}~\bibnamefont{Adam}}, \bibinfo {author}
  {\bibfnamefont{E.~H.}\ \bibnamefont{Hwang}},\ and\ \bibinfo {author}
  {\bibfnamefont{E.}~\bibnamefont{Rossi}},\ }%
  \bibfield{journal}{%
{\bibinfo {journal} {Rev. Mod. Phys.}}\ }%
  \textbf{\bibinfo {volume} {83}},\ \bibinfo {pages} {407} (\bibinfo {year}
  {2011})%
  \bibAnnoteFile{NoStop}{dassarma:2011}%
\bibitem{kim:2011}%
  \BibitemOpen
  \bibfield{author}{%
  \bibinfo {author} {\bibfnamefont{S.}~\bibnamefont{Kim}}, \bibinfo {author}
  {\bibfnamefont{I.}~\bibnamefont{Jo}}, \bibinfo {author}
  {\bibfnamefont{J.}~\bibnamefont{Nah}}, \bibinfo {author}
  {\bibfnamefont{Z.}~\bibnamefont{Yao}}, \bibinfo {author}
  {\bibfnamefont{S.~K.}\ \bibnamefont{Banerjee}},\ and\ \bibinfo {author}
  {\bibfnamefont{E.}~\bibnamefont{Tutuc}},\ }%
  \bibfield{journal}{%
{\bibinfo {journal} {Phys. Rev. B}}\ }%
  \textbf{\bibinfo {volume} {83}},\ \bibinfo {pages} {161401} (\bibinfo {year}
  {2011})%
  \bibAnnoteFile{NoStop}{kim:2011}%
\bibitem{feldman:2009}%
  \BibitemOpen
  \bibfield{author}{%
  \bibinfo {author} {\bibfnamefont{B.~E.}\ \bibnamefont{Feldman}}, \bibinfo
  {author} {\bibfnamefont{J.}~\bibnamefont{Martin}},\ and\ \bibinfo {author}
  {\bibfnamefont{A.}~\bibnamefont{Yacoby}},\ }%
  \bibfield{journal}{%
  \bibinfo {journal} {Nature Phys.}\ }%
  \textbf{\bibinfo {volume} {5}},\ \bibinfo {pages} {889} (\bibinfo {year}
  {2009})%
  \bibAnnoteFile{NoStop}{feldman:2009}%
\bibitem{min:2008}%
  \BibitemOpen
  \bibfield{author}{%
  \bibinfo {author} {\bibfnamefont{H.}~\bibnamefont{Min}}, \bibinfo {author}
  {\bibfnamefont{G.}~\bibnamefont{Borghi}}, \bibinfo {author}
  {\bibfnamefont{M.}~\bibnamefont{Polini}},\ and\ \bibinfo {author}
  {\bibfnamefont{A.~H.}\ \bibnamefont{MacDonald}},\ }%
  \bibfield{journal}{%
{\bibinfo {journal} {Phys. Rev. B}}\ }%
  \textbf{\bibinfo {volume} {77}},\ \bibinfo {pages} {041407} (\bibinfo {year}
  {2008})%
  \bibAnnoteFile{NoStop}{min:2008}%
\bibitem{hwang:2009}%
  \BibitemOpen
  \bibfield{author}{%
  \bibinfo {author} {\bibfnamefont{E.~H.}\ \bibnamefont{Hwang}}\ and\ \bibinfo
  {author} {\bibfnamefont{S.}~\bibnamefont{Das~Sarma}},\ }%
  \bibfield{journal}{%
{\bibinfo {journal} {Phys. Rev. B}}\ }%
  \textbf{\bibinfo {volume} {80}},\ \bibinfo {pages} {205405} (\bibinfo {year}
  {2009})%
  \bibAnnoteFile{NoStop}{hwang:2009}%
\bibitem{nand:2010}%
  \BibitemOpen
  \bibfield{author}{%
  \bibinfo {author} {\bibfnamefont{R.}~\bibnamefont{Nandkishore}}\ and\
  \bibinfo {author} {\bibfnamefont{L.}~\bibnamefont{Levitov}},\ }%
  \bibfield{journal}{%
{\bibinfo {journal} {Phys. Rev. Lett.}}\
  }%
  \textbf{\bibinfo {volume} {104}},\ \bibinfo {pages} {156803} (\bibinfo {year}
  {2010})%
  \bibAnnoteFile{NoStop}{nand:2010}%
\bibitem{gramila:1991}%
  \BibitemOpen
  \bibfield{author}{%
  \bibinfo {author} {\bibfnamefont{T.~J.}\ \bibnamefont{Gramila}}, \bibinfo
  {author} {\bibfnamefont{J.~P.}\ \bibnamefont{Eisenstein}}, \bibinfo {author}
  {\bibfnamefont{A.~H.}\ \bibnamefont{MacDonald}}, \bibinfo {author}
  {\bibfnamefont{L.~N.}\ \bibnamefont{Pfeiffer}},\ and\ \bibinfo {author}
  {\bibfnamefont{K.~W.}\ \bibnamefont{West}},\ }%
  \bibfield{journal}{%
{\bibinfo {journal} {Phys. Rev. Lett.}}\ }%
  \textbf{\bibinfo {volume} {66}},\ \bibinfo {pages} {1216} (\bibinfo {year}
  {1991})%
  \bibAnnoteFile{NoStop}{gramila:1991}%
\bibitem{zheng:1993}
  \BibitemOpen
  \bibfield{author}{%
  \bibinfo {author} {\bibfnamefont{L.}~\bibnamefont{Zheng}}\ and\ \bibinfo
  {author} {\bibfnamefont{A.~H.}\ \bibnamefont{MacDonald}},\ }%
  \bibfield{journal}{%
{\bibinfo {journal} {Phys. Rev. B}}\ }%
  \textbf{\bibinfo {volume} {48}},\ \bibinfo {pages} {8203} (\bibinfo {year}
  {1993})%
  \bibAnnoteFile{NoStop}{zheng:1993}%
\bibitem{flensberg:1995}%
  \BibitemOpen
  \bibfield{author}{%
  \bibinfo {author} {\bibfnamefont{K.}~\bibnamefont{Flensberg}}\ and\ \bibinfo
  {author} {\bibfnamefont{B.~Y.-K.}\ \bibnamefont{Hu}},\ }%
  \bibfield{journal}{%
{\bibinfo {journal} {Phys. Rev. B}}\ }%
  \textbf{\bibinfo {volume} {52}},\ \bibinfo {pages} {14796} (\bibinfo {year}
  {1995})%
  \bibAnnoteFile{NoStop}{flensberg:1995}%
\bibitem{rojo:1999}%
  \BibitemOpen
  \bibfield{author}{%
  \bibinfo {author} {\bibfnamefont{A.~G.}\ \bibnamefont{Rojo}},\ }%
  \bibfield{journal}{%
  \bibinfo {journal} {J. Phys.: Condens. Matter}\ }%
  \textbf{\bibinfo {volume} {11}},\ \bibinfo {pages} {R31} (\bibinfo {year}
  {1999})%
  \bibAnnoteFile{NoStop}{rojo:1999}%
\bibitem{hwang:2003}%
  \BibitemOpen
  \bibfield{author}{%
  \bibinfo {author} {\bibfnamefont{E.~H.}\ \bibnamefont{Hwang}}, \bibinfo
  {author} {\bibfnamefont{S.}~\bibnamefont{Das~Sarma}}, \bibinfo {author}
  {\bibfnamefont{V.}~\bibnamefont{Braude}},\ and\ \bibinfo {author}
  {\bibfnamefont{A.}~\bibnamefont{Stern}},\ }%
  \bibfield{journal}{%
{\bibinfo {journal} {Phys. Rev. Lett.}}\
  }%
  \textbf{\bibinfo {volume} {90}},\ \bibinfo {pages} {086801} (\bibinfo {year}
  {2003})%
  \bibAnnoteFile{NoStop}{hwang:2003}%
\bibitem{dassarma:2005}%
  \BibitemOpen
  \bibfield{author}{%
  \bibinfo {author} {\bibfnamefont{S.}~\bibnamefont{Das~Sarma}}\ and\ \bibinfo
  {author} {\bibfnamefont{E.~H.}\ \bibnamefont{Hwang}},\ }%
  \bibfield{journal}{%
{\bibinfo {journal} {Phys. Rev. B}}\ }%
  \textbf{\bibinfo {volume} {71}},\ \bibinfo {pages} {195322} (\bibinfo {year}
  {2005})%
  \bibAnnoteFile{NoStop}{dassarma:2005}%
\bibitem{tse:2007}%
  \BibitemOpen
  \bibfield{author}{%
  \bibinfo {author} {\bibfnamefont{W.-K.}\ \bibnamefont{Tse}}, \bibinfo
  {author} {\bibfnamefont{B.~Y.-K.}\ \bibnamefont{Hu}},\ and\ \bibinfo {author}
  {\bibfnamefont{S.}~\bibnamefont{Das~Sarma}},\ }%
  \bibfield{journal}{%
{\bibinfo {journal} {Phys. Rev. B}}\ }%
  \textbf{\bibinfo {volume} {76}},\ \bibinfo {pages} {081401} (\bibinfo {year}
  {2007})%
  \bibAnnoteFile{NoStop}{tse:2007}%
\bibitem{novoselov:2005}%
  \BibitemOpen
  \bibfield{author}{%
  \bibinfo {author} {\bibfnamefont{K.~S.}\ \bibnamefont{Novoselov~{\it et
  al.}}},\ }%
  \bibfield{journal}{%
  \bibinfo {journal} {Nature}\ }%
  \textbf{\bibinfo {volume} {438}},\ \bibinfo {pages} {197} (\bibinfo {year}
  {2005})%
  \bibAnnoteFile{NoStop}{novoselov:2005}%
\bibitem{tan:2007}%
  \BibitemOpen
  \bibfield{author}{%
  \bibinfo {author} {\bibfnamefont{Y.-W.}\ \bibnamefont{Tan}}, \bibinfo
  {author} {\bibfnamefont{Y.}~\bibnamefont{Zhang}}, \bibinfo {author}
  {\bibfnamefont{K.}~\bibnamefont{Bolotin}}, \bibinfo {author}
  {\bibfnamefont{Y.}~\bibnamefont{Zhao}}, \bibinfo {author}
  {\bibfnamefont{S.}~\bibnamefont{Adam}}, \bibinfo {author}
  {\bibfnamefont{E.~H.}\ \bibnamefont{Hwang}}, \bibinfo {author}
  {\bibfnamefont{S.}~\bibnamefont{Das~Sarma}}, \bibinfo {author}
  {\bibfnamefont{H.~L.}\ \bibnamefont{Stormer}},\ and\ \bibinfo {author}
  {\bibfnamefont{P.}~\bibnamefont{Kim}},\ }%
  \bibfield{journal}{%
{\bibinfo {journal} {Phys. Rev. Lett.}}\
  }%
  \textbf{\bibinfo {volume} {99}},\ \bibinfo {pages} {246803} (\bibinfo {year}
  {2007})%
  \bibAnnoteFile{NoStop}{tan:2007}%
\bibitem{chen:2008}%
  \BibitemOpen
  \bibfield{author}{%
  \bibinfo {author} {\bibfnamefont{J.~H.}\ \bibnamefont{Chen}}, \bibinfo
  {author} {\bibfnamefont{C.}~\bibnamefont{Jang}}, \bibinfo {author}
  {\bibfnamefont{M.~S.}\ \bibnamefont{Fuhrer}}, \bibinfo {author}
  {\bibfnamefont{E.~D.}\ \bibnamefont{Williams}},\ and\ \bibinfo {author}
  {\bibfnamefont{M.}~\bibnamefont{Ishigami}},\ }%
  \bibfield{journal}{%
  \bibinfo {journal} {Nature Phys.}\ }%
  \textbf{\bibinfo {volume} {4}},\ \bibinfo {pages} {377} (\bibinfo {year}
  {2008})%
  \bibAnnoteFile{NoStop}{chen:2008}%
\bibitem{flensberg:1995a}%
  \BibitemOpen
  \bibfield{author}{%
  \bibinfo {author} {\bibfnamefont{K.}~\bibnamefont{Flensberg}}, \bibinfo
  {author} {\bibfnamefont{B.~Y.-K.}\ \bibnamefont{Hu}}, \bibinfo {author}
  {\bibfnamefont{A.-P.}\ \bibnamefont{Jauho}},\ and\ \bibinfo {author}
  {\bibfnamefont{J.~M.}\ \bibnamefont{Kinaret}},\ }%
  \bibfield{journal}{%
{\bibinfo {journal} {Phys. Rev. B}}\ }%
  \textbf{\bibinfo {volume} {52}},\ \bibinfo {pages} {14761} (\bibinfo {year}
  {1995})%
  \bibAnnoteFile{NoStop}{flensberg:1995a}%
\bibitem{kamenev:1995}%
  \BibitemOpen
  \bibfield{author}{%
  \bibinfo {author} {\bibfnamefont{A.}~\bibnamefont{Kamenev}}\ and\ \bibinfo
  {author} {\bibfnamefont{Y.}~\bibnamefont{Oreg}},\ }%
  \bibfield{journal}{%
{\bibinfo {journal} {Phys. Rev. B}}\ }%
  \textbf{\bibinfo {volume} {52}},\ \bibinfo {pages} {7516} (\bibinfo {year}
  {1995})%
  \bibAnnoteFile{NoStop}{kamenev:1995}%
\bibitem{tse:2007a}%
  \BibitemOpen
  \bibfield{author}{%
  \bibinfo {author} {\bibfnamefont{W.-K.}\ \bibnamefont{Tse}}\ and\ \bibinfo
  {author} {\bibfnamefont{S.}~\bibnamefont{Das~Sarma}},\ }%
  \bibfield{journal}{%
{\bibinfo {journal} {Phys. Rev. B}}\ }%
  \textbf{\bibinfo {volume} {75}},\ \bibinfo {pages} {045333} (\bibinfo {year}
  {2007})%
  \bibAnnoteFile{NoStop}{tse:2007a}%
\bibitem{hwang:2007}%
  \BibitemOpen
  \bibfield{author}{%
  \bibinfo {author} {\bibfnamefont{E.~H.}\ \bibnamefont{Hwang}}\ and\ \bibinfo
  {author} {\bibfnamefont{S.}~\bibnamefont{Das~Sarma}},\ }%
  \bibfield{journal}{%
{\bibinfo {journal} {Phys. Rev. B}}\ }%
  \textbf{\bibinfo {volume} {75}},\ \bibinfo {pages} {205418} (\bibinfo {year}
  {2007})%
  \bibAnnoteFile{NoStop}{hwang:2007}%
\bibitem{sensarma:2010}%
  \BibitemOpen
  \bibfield{author}{%
  \bibinfo {author} {\bibfnamefont{R.}~\bibnamefont{Sensarma}}, \bibinfo
  {author} {\bibfnamefont{E.~H.}\ \bibnamefont{Hwang}},\ and\ \bibinfo {author}
  {\bibfnamefont{S.}~\bibnamefont{Das~Sarma}},\ }%
  \bibfield{journal}{%
{\bibinfo {journal} {Phys. Rev. B}}\ }%
  \textbf{\bibinfo {volume} {82}},\ \bibinfo {pages} {195428} (\bibinfo {year}
  {2010})%
  \bibAnnoteFile{NoStop}{sensarma:2010}%
\bibitem{hwang_sds:drag}%
  \BibitemOpen
  \bibfield{author}{%
  \bibinfo {author} {\bibfnamefont{E.~H.}\ \bibnamefont{Hwang}}\ and\ \bibinfo
  {author} {\bibfnamefont{S.~D.}\ \bibnamefont{Sarma}},\ }%
  \bibfield{journal}{%
  \bibinfo {journal} {arXiv:1105.3203}}%
   (\bibinfo {year} {2011})%
  \bibAnnoteFile{NoStop}{hwang_sds:drag}%
\bibitem{narozhny:2007}%
  \BibitemOpen
  \bibfield{author}{%
  \bibinfo {author} {\bibfnamefont{B.~N.}\ \bibnamefont{Narozhny}},\ }%
  \bibfield{journal}{%
{\bibinfo {journal} {Phys. Rev. B}}\ }%
  \textbf{\bibinfo {volume} {76}},\ \bibinfo {pages} {153409} (\bibinfo {year}
  {2007})%
  \bibAnnoteFile{NoStop}{narozhny:2007}%
\bibitem{peres:2011}%
  \BibitemOpen
  \bibfield{author}{%
  \bibinfo {author} {\bibfnamefont{N.~M.~R.}\ \bibnamefont{Peres}}, \bibinfo
  {author} {\bibfnamefont{J.~M.~B.}\ \bibnamefont{{Lopes dos Santos}}},\ and\
  \bibinfo {author} {\bibfnamefont{A.~H.}\ \bibnamefont{{Castro Neto}}},\ }%
  \bibfield{journal}{%
  \bibinfo {journal} {arXiv:1105.5399; Europhys. Lett.}\ }%
  \textbf{\bibinfo {volume} {95}},\ \bibinfo {pages} {18001} (\bibinfo {year}
  {2011})%
  \bibAnnoteFile{NoStop}{peres:2011}%
\bibitem{katsnelson:2011}%
  \BibitemOpen
  \bibfield{author}{%
  \bibinfo {author} {\bibfnamefont{M.~I.}\ \bibnamefont{Katsnelson}},\ }%
  \bibfield{journal}{%
{\bibinfo {journal} {arXiv:1105.2534; Phys.
  Rev. B}}\ }%
  \textbf{\bibinfo {volume} {84}},\ \bibinfo {pages} {041407} (\bibinfo {year}
  {2011})%
  \bibAnnoteFile{NoStop}{katsnelson:2011}%
\end{thebibliography}

%

\end{document}